\documentclass[prd,twocolumn,nofootinbib,amsfonts]{revtex4-2}
\usepackage[utf8]{inputenc}
\usepackage[T1]{fontenc}
\usepackage[colorlinks, linkcolor = red]{hyperref}
\usepackage{graphicx,bm,amsmath,scalerel}
\usepackage{bbold}
\usepackage{wasysym}
\usepackage{verbatim}
\usepackage{amssymb}
\usepackage{epstopdf}
\usepackage{xmpmulti}
\usepackage{centernot}
\usepackage{multirow}
\usepackage{tikz}
\usepackage{graphicx}
\usepackage{float}
\usepackage{mathtools}
\usepackage{comment}

\newcommand{\dd}{{\rm d}}

\makeatletter

\newcommand{\be}{\begin{equation}}
	\newcommand{\ee}{\end{equation}}
\newcommand{\beq}{\begin{eqnarray}}
	\newcommand{\eeq}{\end{eqnarray}}
\newcommand{\ba}{\begin{align}}
	\newcommand{\ea}{\end{align}}

\begin{document}

    \title{Thick accretion disk configurations in the Born-Infeld teleparallel gravity}

	\author{Sebastian Bahamonde}
    \email{bahamonde.s.aa@m.titech.ac.jp}   \affiliation{Department of Physics, Tokyo Institute of Technology 1-12-1 Ookayama, Meguro-ku, Tokyo 152-8551 Japan,}
    \affiliation{Laboratory of Theoretical Physics, Institute of Physics, University of Tartu, W. Ostwaldi 1, 50411 Tartu, Estonia}
    
    \author{Shokoufe Faraji}
    \email{shokoufe.faraji@zarm.uni-bremen.de}
	\affiliation{University of Bremen, Center of Applied Space Technology and Microgravity (ZARM), 28359 Bremen, Germany}

    \author{Eva Hackmann}
    \email{eva.hackmann@zarm.uni-bremen.de}
	\affiliation{University of Bremen, Center of Applied Space Technology and Microgravity (ZARM), 28359 Bremen, Germany}
	
	\author{Christian Pfeifer}
	\email{christian.pfeifer@zarm.uni-bremen.de}
	\affiliation{University of Bremen, Center of Applied Space Technology and Microgravity (ZARM), 28359 Bremen, Germany}

\begin{abstract}
The main goal of this paper is to investigate one of the important astrophysical systems, namely Thick accretion disks, in the background of the spherically symmetric solution in Born-Infeld teleparallel gravity to examine 
observable predictions of the theory in the vicinity of black holes.
Thus, the properties of the non-self-gravitating equilibrium surfaces characterising the Thick accretion disks model are studied.
In addition, we find an observational bound on the parameter of the model as $\lambda\gtrsim 140$. We show this analytical accretion disk model for different values of $\lambda$ and compare the result with the corresponding Schwarzschild solution in the general theory of relativity.

\end{abstract}

\maketitle


\section{Introduction}
Accretion disks are assumed to be responsible for high-energy astronomical observations since they can reach deep into the strong gravity regime of black holes and compact objects. Therefore, their properties offer the opportunity to test predictions from our understanding of the gravitational interaction based on the general relativity (GR) or its numerous modifications and extensions in an environment which is otherwise hard to access. Hence, the theory of accretion disks of black holes belongs to the area of fundamental physics.

In this area, the Thick accretion disk is an analytical model that analyse the shape of equipotential surfaces in a given background. This geometrically thick configuration originated in the seminal papers \cite{1974AcA....24...45A,1978A&A....63..209K,1980AcA....30....1J,1980A&A....88...23P,1980ApJ...242..772A,1982MitAG..57...27P,1982ApJ...253..897P}. In fact, the Thick disk model is believed to exist in the vicinity of X-ray binaries, active galactic nuclei, and in the central engine of gamma-ray bursts. In general, such a disk has a toroidal shape, large optical depth, optically thick, super-Eddington accretion, radiation pressure support, and is highly radiatively inefficient. Of course, there are various studies on the accretion disks in the scope of GR (for a review see e.g. \cite{2013LRR....16....1A}) and a few in generalized theories \cite{2019PhRvD..99d3002G,2021PhRvD.103l4009C,2021PhRvD.104j3008G,2022EPJC...82..190K}. 

Changing the underlying geometry describing gravity is one possible route to modify GR. Instead of a metric and its Levi-Civita connection, one could employ a metric and an arbitrary connection, as is the case in the metric affine theories of gravity or more specific connections with specific properties. Such theories are then often tested within cosmology \cite{CANTATA:2021ktz} or investigated as low energy-limit of quantum gravity \cite{Addazi:2021xuf}. Among the most famous models suggested in the literature are teleparallel gravity theories~\cite{Aldrovandi:2013wha, Krssak:2015oua, Bahamonde:2021gfp}. In this case, one use a flat and metric compatible connection with torsion. When the metric is replaced by its tetrads as fundamental variable, teleparallel gravity can also be understood as an approach to formulate gravity as a gauge theory~\cite{Pereira:2019woq,Huguet:2020ler}.

It is possible to formulate GR in terms of the teleparallel geometry variables, which is known for long as the "teleparallel equivalent of GR" (TEGR)~\cite{Maluf:2013gaa,Garecki:2010jj,BeltranJimenez:2019tjy}. This theory is constructed from a specific sum of contractions of the torsion tensor, known as the torsion scalar~$T$. It turns out that TEGR has the exact predictions as GR, while all the gravitational effects can be understood as originating from torsion. Starting from TEGR, extensions and modifications have been formulated which do not have an equivalent in modifications of GR based on semi-Riemannian geometry. A particular advantage of constructing modified theories of gravity based on teleparallel gravity is the absence of a Lovelock theorem due to the structure of the torsion tensor as a fundamental building block for the theories instead of the curvature. One of the most popular modifications of TEGR is $f(T)$-gravity which is constructed using a function $f$ of the torsion scalar $T$ in the action defining the theory~\cite{Ferraro:2006jd,Ferraro:2008ey,Chen:2010va,Cai:2015emx,Golovnev:2020zpv}, in analogy to $f(R)$-theories of gravity. For the interested reader on the formulation of this theory and its application, see the review~\cite{Bahamonde:2021gfp}. 

For general $f(T)$-theories, it is not easy to obtain solutions to the gravitational field equations, even with further symmetry assumptions. However, the recent detailed studies about spherical symmetry in $f(T)$-gravity~\cite{DeBenedictis:2016aze,Bahamonde:2019zea,Bahamonde:2020vpb,Pfeifer:2021njm,DeBenedictis:2022sja} led successfully to a non-perturbative analytic spherically symmetric black hole solution in Born-Infeld $f(T)$-gravity~\cite{Bahamonde:2021srr}; the latter being a teleparallel gravity theory with a specific choice of the function $f$ which resembles Born-Infeld electrodynamics. Born-Infeld $f(T)$-gravity is a particularly interesting model since it is capable to initiate inflation without the need of introducing an inflation~\cite{Ferraro:2006jd,Ferraro:2008ey} and contains (at least on the perturbative level) regularized (Ba\~nados, Teitelboim, Zanelli) BTZ black hole solutions~\cite{Bohmer:2020jrh}.

It is worth mentioning that on the phenomenological side $f(T)$-gravity offers promising modifications of general relativity, for example to explain the dark energy, to build viable cosmological models and to avoid singularities, e.g.  \cite{2021JCAP...07..052H,2022arXiv220700059A}. However, on the mathematical foundational side, there are still open questions. For instance, $f(T)$-gravity predicts strongly coupled gravitational degrees of freedom around linear Minkowski and cosmological perturbations, e.g. \cite{2020arXiv200407536B,2021PhRvD.103d4009G,Bahamonde:2022ohm,Golovnev:2020aon,Golovnev:2018wbh}. Their physical impacts are under debate. Also, the role of different realizations of (local) Lorentz invariance, with and without taking the spin connection into account \cite{Krssak:2015oua,2021CQGra..38s7001G,Golovnev:2020zpv}, and the number of propagating degrees of freedom \cite{Blixt:2020ekl,Ferraro:2018axk,BeltranJimenez:2020fvy,Golovnev:2020nln,Blagojevic:2020dyq}, are still under study.


In this work, we explore the properties of the equilibrium configuration of an accretion disk in the background of the Born-Infeld teleparallel gravity generalization of Schwarzschild geometry. This study is the next step in investigating the viability of this solution.

The structure of this paper is as follows. In Section~\ref{sec:BITP} we briefly recall necessary notions about teleparallel gravity in general, $f(T)$-gravity as well as particularly Born-Infeld gravity and the spherically symmetric solution which is the background geometry for the analysis of the accretion disk. Afterwards, in Section~\ref{sec:PM} we study point particle motion in this metric. The Thick disk model is briefly described in Section~\ref{accretionmodel}. The structure of the disk in this background is presented in Section \ref{sec:RandD}. We summarize our results and conclude in Section~\ref{sum}.

The notational conventions in this article are  $(-,+,+,+)$ for the signature of the metric and geometrical units with $c=G=1$; in addition in the result section we also assume $M=1$. In addition the "over-dot" is used for the derivation with respect to the affine parameter, and the "prime" for the derivative with respect to coordinate~$r$.

\section{Born-Infeld teleparallel gravity}\label{sec:BITP}
We briefly summarize the path to the analytic spherically symmetric black hole solution presented in \cite{Bahamonde:2021srr} of Born-Infeld teleparallel gravity \cite{Ferraro:2006jd}. The basic notions of teleparallel gravity and $f(T)$-gravity are recalled in this section. We follow the standard references for teleparallel gravity \cite{Aldrovandi:2013wha,Hohmann:2017duq,Krssak:2015oua,Bahamonde:2021gfp}, where more details can be found.

\subsection{Theory and field equations}
The fundamental fields in teleleparallel gravity are a tetrad $\theta^a = \theta^a{}_\mu dx^\mu$, which determines the spacetime metric via
\begin{align}\label{eq:met}
    g_{\mu\nu} = \eta_{ab}\theta^a{}_\mu\theta^b{}_\nu\,,
\end{align}
where $\eta_{ab}\sim\textrm{diag}(-,+,+,+)$ is the Minkowski metric, and a flat metric compatible spin connection $\omega^a{}_{b\mu}$ with torsion. The flatness and metric compatibility condition imply that the spin connection coefficients are generated by local Lorentz matrices $\Lambda^a{}_b$
\begin{align}
    \omega^a{}_{b\mu} = \Lambda^a{}_c\partial_\mu (\Lambda^{-1})^c{}_b\,,
\end{align}
while the torsion tensor components are given by
\begin{equation}
T^{\rho}{}_{\mu\nu} = e_a{}^{\rho}\left(\partial_{\mu}e^a{}_{\nu} - \partial_{\nu}e^a{}_{\mu} + \omega^a{}_{b\mu}e^b{}_{\nu} - \omega^a{}_{b\nu}e^b{}_{\mu} \right)\,.
\end{equation}
Due to the expression of the spin connection coefficients in terms of the Lorentz matrices one can always fix a Lorentz frame, the so-called Weitzenb\"ock gauge, such that the spin connection coefficients are zero, and perform all derivations in this frame, which is what is usually done in teleparallel gravity.

Teleparallel theories of gravity are derived from an action which is built from the scalars constructed from the torsion tensor. One of these, called the torsion scalar, is of particular interest since it defines TEGR, namely,
\begin{align}
    T=\frac{1}{4}T^{\mu\nu\rho}T_{\mu\nu\rho} + \frac{1}{2}T^{\mu\nu\rho}T_{\rho\nu\mu} - T_{\rho}T^{\rho}\,\label{defT}.
\end{align}
Employing the action
\begin{align}\label{eq:tegract}
    S = \frac{1}{2\kappa^2}\int_M  \dd^4x |\theta| T \,,
\end{align}
where $\kappa^2 = 8\pi G/c^4$ is the gravitational constant, and $|\theta|$ is the norm of the determinant of the tetrad, yields field equations for the tetrad, whose solutions are tetrads whose metric is a solution of the Einstein vacuum equations. Matter coupling can be added, but since we consider vacuum solutions in this paper we do not discuss the details of matter coupling here \footnote{A detailed discussion on the matter coupling in teleparallel gravity can be found in \cite{BeltranJimenez:2020sih}, references therein, and in the reviews and overview articles mentioned in the beginning of this section.}.

To address questions like: dark matter, dark energy and further shortcomings between the predictions of general relativity and observations, extension of TEGR have been suggest. In analogy to $f(R)$-gravity, one of the most studied is $f(T)$-gravity. It is defined by the action
\begin{align}
    S = \frac{1}{2\kappa^2}\int_M \dd^4x |\theta| f(T)\,,
\end{align}
where $f$ is an, in principle, arbitrary function of $T$. The vacuum field equations of this theory can be expressed as
\begin{equation}
f_{T}G_{\mu\nu}+\frac12 \left(f-f_{T}T\right)g_{\mu\nu}+S_{\mu\nu\alpha}\partial^{\alpha}f_T=0\,, \label{eq:fT}
\end{equation}
where $f_T=df/dT$ and $G_{\mu\nu}$ is the Einstein tensor derived from the Levi-Civita connection of the metric \eqref{eq:met}. This form of the field equation is very convenient and it also clearly shows that for $f(T)=T$, they reduce to~$G_{\mu\nu}=0$. 



A particularly interesting choice of $f$, which goes under the name of teleparallel Born-Infeld gravity, is 
\begin{equation}\label{eq:BIact}
    f(T)=\hat \lambda\Big(\sqrt{1+\frac{2 T}{\hat \lambda}}-1\Big)\,.
\end{equation}
It has been shown that this theory explains inflation purely geometrically \cite{Ferraro:2006jd}, and contains non-singular black holes \cite{Bohmer:2020jrh}. For $\hat \lambda \to \infty$ the theory becomes TEGR (and thus GR).

\subsection{Analytic spherically symmetric solution}

The teleparallel description is generally given in terms of tetrads instead of the metric itself. In $f(T)$-gravity in spherical symmetry, after assuming that not only the tetrad respects spherical symmetry but also the teleparallel connection, two branches of solutions exist for the anti-symmetric part of the field equations~\eqref{eq:fT}, one based on real tetrads, and another on complex tetrads, see \cite{Bahamonde:2021srr} for all details. 

The solution we discuss here has been obtained in the complex tetrads branch. Such a complex tetrad cannot immediately be interpreted as an observer frame, but only be used as description of the gravitational degrees of freedom. It does not cause any problem, as long as it is ensured that all physical observables, and in particular the metric, are real. For gravitational perturbation theory this latter condition poses a constraint on the allowed complex perturbations. Since we do not study perturbations in the following, but properties of a non-perturbative solution of the field equations, we do not analyse these conditions here, and leave them for future research. Also, so far no non-perturbative solution could be found in $f(T)$-gravity on the basis of the real tetrad.

Born-Infeld teleparallel gravity, defined by \eqref{eq:BIact}, has the advantage that an analytical solution of the field equations \eqref{eq:fT} in spherical symmetry could be found based on a complex tetrads \cite{Bahamonde:2021srr}, which can be interpreted as generalization of Schwarzschild solution. The resulting metric is

\begin{align}\label{eq:com_f_T_metric}
    \dd s^2&=-\frac{a_1^2 }{r}S \dd t^2+\frac{\hat \lambda ^{5/2} r^5}{(4 + r^2 \hat \lambda)^2}S^{-1}\dd r^2+r^2\dd\Omega^2\,,
\end{align}
where $\dd\Omega^2=\dd\theta^2+\sin^2\theta\, \dd\phi^2$ and
\begin{align}
    S:= \sqrt{\hat \lambda } (a_0 \hat{\lambda} +r)-2 \tan ^{-1}\left(\frac{\sqrt{\hat \lambda } r}{2}\right).
\end{align}
It contains two constants of integration $a_0$ and $a_1$ as well as the theory parameter $\hat \lambda$. In order to have asymptotically flat and the proper Schwarzschild limit for $\hat \lambda\to \infty$ we need to specify $a_0$ and $a_1$. Further, for simplicity reason, we introduce the dimensionless quantity

\begin{align}
    \lambda = M \sqrt{\hat \lambda}.
\end{align}
Therefore, we have

\begin{align}
    a_0=-\frac{2M^3}{\lambda^2} \quad \text{,} \quad a_1=\sqrt{\frac{M}{\lambda}}.
\end{align}
Finally, the metric becomes
\begin{align}\label{eq:fTBImetric}
    \dd s^2 = - A(r) \dd t^2 +  \frac{B(r)}{A(r)}\dd r^2 + r^2\dd \Omega^2\,,
\end{align}
where 
\begin{align}
    A(r) & :=1-\frac{2 M}{r}-\frac{2 M}{r \lambda} \mathcal{T}\,, \label{defA}\\
    B(r) & :=\frac{r^4 \lambda^4}{16 M^4 \left(1+\frac{\lambda ^2 r^2}{4 M^2}\right)^2}\,,
\end{align}
and 

\begin{align}
\mathcal{T}=\tan ^{-1}\left(\frac{\lambda  r}{2 M}\right).    
\end{align}
To gain a better insight into the physical interpretation of the parameter $\lambda$, we derive the Komar mass $\mathcal{M}$ for the metric. It is a measure of the force needed by an observer at infinity to keep a spherical uniform mass distribution in place and relies on the existence of a timelike killing vector field, and that matter follows the autoparallels of the Levi-Civita connection of the metric \cite{Wald}. We find
\begin{align}
    \mathcal{M} &= 
    \frac{1}{2}\lim_{r\to\infty}\left( r^2 \frac{g_{tt}'}{g_{tt}} \right) = \left(\frac{\pi }{2\lambda}+1\right) M\,.\
\end{align}
In teleparallel gravity, the expression and interpretation of the Komar mass stay the same for stationary asymptotically flat spacetimes, as long as matter follows the geodesics of the Levi-Civita connection of the metric and is not influenced by torsion, which is what we assume, as already discussed below \eqref{eq:tegract}.

Expanding the metric for large $\frac{r}{M}$  to study the weak field limit gives
\begin{align}
  - g_{tt} & = 1 - \frac{2M}{r} \left( 1 + \frac{\pi}{2\lambda} \right) + \frac{4M^2}{r^2 \lambda^2} + \mathcal{O}(r^{-4})\,,\\
    g_{rr} & = 1 + \frac{2M}{r} \left( 1 + \frac{\pi}{2\lambda} \right) + \mathcal{O}(r^{-2}).
\end{align}
Comparing this expression to the standard PPN expansion
\begin{align}
   - g_{tt} & = 1 - \frac{2\hat{M}}{r} + (\beta-\gamma) \frac{2\hat{M}^2}{r^2}\\
     g_{rr} & = 1 + \gamma \frac{2\hat{M}}{r},
\end{align}
where $\hat{M} = Gm/c^2$ and $m$ is the Newtonian mass, we find consistently that $\hat{M}=\mathcal{M}$. Thus, $\gamma=1$ agrees with its GR value; however
\begin{align}
    (\beta-1) &= \frac{8}{\pi^2} \Big(1-\frac{M}{\hat{M}}\Big)^2 = \frac{8}{(2\lambda+\pi)^2},
\end{align}
attains a correction. Using the observational bound that $|\beta-1|<10^{-4}$~\cite{Will:2014kxa}, we find $\lambda\gtrsim 140$.

Nevertheless, this finding deserves an additional comment. In the $f(T)$-gravity literature, it has been shown that the PPN parameters of the theory coincide with the ones from GR~\cite{Ualikhanova:2019ygl}. The source for this mismatch is related to the assumptions. In fact, in the mentioned literature, the computation is based on the Minkowski background to have tetrad in Cartesian coordinates as $e^a{}_\mu=(1,1,1,1)$. However, the tetrad we employed in the Minkowski limit has a different form in Cartesian coordinates. This means that the PPN analysis in the previous works does not contain the complex tetrad in the Minkowski limit and does not apply here.

In the following, we will switch to dimensionless values by using $\tilde{r}=r/M$ but dropping the twidle for notational convenience. This is equivalent to choosing $M=1$.


\section{Particle motion}\label{sec:PM}
Before we study the accretion disk in the next section, we first derive the radius of the marginally stable orbit $r_{\textrm{ms}}$ and the marginally bound orbit $r_{\textrm{mb}}$ obtain from the metric \eqref{eq:fTBImetric}, as two essential quantities for determining the Thick disk model in a given spacetime. The Lagrangian is given by

\begin{align}
    \mathcal{L} = \frac{1}{2} g_{\mu\nu}(x)\dot x^\mu \dot x^\nu = -\frac{\sigma}{2},
\end{align}
Of course, by this convention $\sigma = 1$ for massive and $\sigma= 0$ for massless particles. Using the conserved quantities $E = \partial_{\dot t} \mathcal{L}$ and $L = \partial_{\dot \phi} \mathcal{L}$ we find
\begin{align}
    \dot r^2 = \frac{1}{B} \left(E^2 - V_{\rm eff}(r)\right)\,,
\end{align}
where

\begin{align}
    V_{\rm eff} & = \frac{(\sigma r^2 + L^2)A}{r^2}\,.
\end{align}
A power series expansion around $\lambda \to \infty$ gives consistently to lowest order the effective potential which determines particle motion in Schwarzschild geometry.

In what follows, we calculate the radii of the marginally stable orbit $r_\textrm{ms}$ (also known as Innermost Stable Circular Orbit "ISCO") and the marginally bound orbit $r_{\textrm{mb}}$. They imply restrictions on the chosen angular momentum distribution in constructing the Thick disk model (see Section \ref{accretionmodel}).


To study the $r_\textrm{ms}$ for massive particles we use the standard procedure (see e.g. \cite{Berry:2020ntz,Song:2021ziq}). For circular orbits, of course we have that $r=\textrm{const.}$, i.e. $\dot{r}=0$, $\ddot{r}=0$. These conditions are to
\begin{enumerate}
\item $V_{\textrm{eff}}(r)=E^2$,
\item $V'_{\textrm{eff}}(r)=0$.
\end{enumerate} 
In addition, to have a stable circular orbit $V''_{\textrm{eff}}(r)>0$ must hold. This states that the circular orbit should lie in a minimum of the potential. 

Therefore, the marginally stable radius $r_\textrm{ms}$ is characterized by $V''_{\textrm{eff}}(r_\textrm{ms})=0$ in addition to the circularity conditions above. Considering the first derivative of the potential as a function of $r$ and $L$ this is equivalent to $L'(r_\textrm{ms}) = 0$.

\begin{figure}
    \centering
    \includegraphics[width=0.9\hsize]{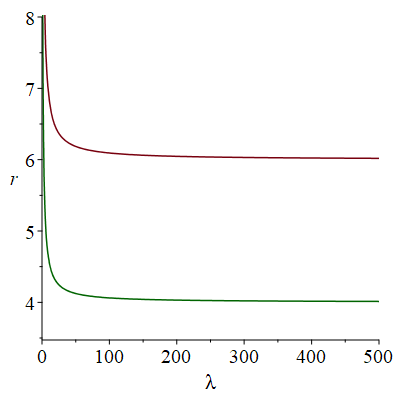}
    \caption{The marginally stable orbit $r_{\rm ms}$ is plotted in red and the marginally bound orbit $r_{\rm mb}$ in green, as functions of the spacetime parameter $\lambda$.}
    \label{fig1}
\end{figure}

From the two circularity conditions above, one can generally solve for the energy and angular momentum parameters
\begin{align}
    E^2 & = \frac{2A^2}{2A-A'r}\,,\\
    L^2 & = \frac{A'r^3}{2A-A'r}\,.
\end{align}
Inserting the metric function $A$ specified in \eqref{defA} (remembering we chose $M=1$) we find from this
\begin{align}
    E^2 & = \frac{(4+r^2\lambda^2)(\lambda(r-2)-2\mathcal{T})^2}{r\lambda\big[r^2(r-3)\lambda^3+6(r-2)\lambda-3(4+r^2\lambda^2)\mathcal{T}\big]}\,, \label{eq:Esq}\\
    L^2 & = \frac{r^2\big[r^2\lambda^3-2(r-2)\lambda+(4+r^2\lambda^2)\mathcal{T}\big]}{r^2(r-3)\lambda^3+6(r-2)\lambda-3(4+r^2\lambda^2)\mathcal{T}}\,.
\end{align}
Note that both $E^2$ and $L^2$ diverge at the common root of their denominators, that is approximately given as
\begin{align}\label{rph}
    r & = 3 + \frac{3\pi}{2\lambda},
\end{align}
to first order in $\lambda^{-1}$, and corresponds to the photon sphere.



Finally, using the expressions for $E^2$ and $L^2$ above, $V''_{\textrm{eff}}(r_\textrm{ms})=0$ is equivalent to
\begin{widetext}
\begin{align}
     & \frac{4r^2\lambda^2(12-r^2\lambda^2)-2r\lambda(12+r^2\lambda^2)(8+r^2\lambda^2)(\mathcal{T}+\lambda)+12(4+r^2\lambda^2)^2(\mathcal{T}+\lambda)^2}{r^3\lambda(4+r^2\lambda^2)\big[r^2(r-3)\lambda^3+6(r-2)\lambda-3(4+r^2\lambda^2)\mathcal{T}\big]}\Big{|}_{r_\textrm{ms}}=0\label{eq:ISCO}\,.
\end{align}
\end{widetext}
To the first order in $\lambda^{-1}$, this equation is solved by
\begin{align}\label{rms}
    r_{\rm ms} & = 6 + \frac{3\pi}{\lambda}\,.
\end{align}

We solve equation \eqref{eq:ISCO} numerically for different values of $\lambda$ in Table \ref{TTb}. A plot of $r_{\rm ms}$ as a function of $\lambda$ is shown in Fig.~\ref{fig1}.

In addition, the marginally bound orbit $r_{\textrm{mb}}$ is the innermost unstable circular orbit, from where infinity can just be reached. It is determined by solving $V_{\textrm{eff}}(r_{\textrm{mb}}) = \sigma$ and $V'_{\textrm{eff}}(r_{\textrm{mb}}) = 0$ with $\sigma=1$ . This corresponds to setting $E^2=1$ in equation \eqref{eq:Esq} 

\begin{align}\label{eq:IBSO}
    &\left(4 +\lambda ^2 r^2\right) \mathcal{T} \left(8 \lambda  +4  \mathcal{T}-\lambda  r\right)\nonumber\\
    &+\lambda ^4 r^2 (4 -r)+2 \lambda ^2  (2 -r) (4 +r)\big{|}_{r_{\rm mb}} = 0\,.
\end{align}
To first order in $\lambda^{-1}$ this equation is solved by
\begin{align}\label{rmb}
    r_{\rm mb} & = 4 + \frac{2\pi}{\lambda}\,.
\end{align}
As for the $r_\textrm{ms}$, we solve \eqref{eq:IBSO} numerically and display the solutions for different values of $\lambda$ in Table~\ref{TTb}.

To summarize the results of this section, we display the $r_\textrm{ms}$ and the $r_{\textrm mb}$ as function of $\lambda$ in Fig.~\ref{fig1}. We see the difference is mostly for smaller values of $\lambda$ and the area between them becomes larger. Therefore, from the construction of the Thick disk model we expected to have a larger disk structure for smaller values of $\lambda$.

\begin{table}
\begin{center} 
\begin{tabular}{c|c|c|c|c|c|c|c|c|c|c|c|c|} 
$\lambda$   & 0.01 & 1 & 10 & 50 & 100 & 500 \\ \hline
$r_{\textrm{mb}}$ & 159.387 & 8.55449 & 4.5936 & 4.12411 & 4.06244 & 4.01255  \\
\hline
$r_{\textrm{ms}}$ & 253.513 & 12.8391 & 6.8904 & 6.18617 & 6.09366 & 6.00942 \\
\end{tabular}
\end{center}
\caption{\label{TTb} Explicit numerical values for the marginally bound $r_{\rm mb}$ and marginally stable $r_{\rm ms}$ orbit in dependence of $\lambda$.}
\end{table}



\section{Thick accretion disk model} \label{accretionmodel}

In this work, we consider the Thick disk model based on the constant angular momentum and the perfect fluid energy-momentum tensor. In addition, it is assumed that the influence of the disk on the underlying spacetime is negligible. The Thick disk model is a description of an accretion disk which is governed by a strong gravitational field and the pressure within the perfect fluid in the fixed specified background. It takes advantage of the fact that the boundary of a barotropic perfect fluid and stationary body is an equipotential surface. One of the immediate outcomes of this hydrodynamical structure is to neglect the accretion flow. Moreover, this model is axisymmetric and stationary; accordingly, the physical and geometrical quantities only depend on $r$ and $\theta$ coordinates Furthermore, the rotation of perfect fluid is assumed to be in the azimuthal direction. For a review of the Thick disk model we refer to \cite{2013LRR....16....1A} and references therein. In this model, the four-velocity and stress-energy tensor reduce to
\begin{align}
    u^{\mu}&=(u^t,0,0,u^{\phi})\,,\\
    T^{\mu}{}_{\nu}&=(\epsilon +p)u_{\nu}u^{\mu} + \delta^{\mu}{}_{\nu} p\,,
\end{align}
where $u^\mu$ is the 4-velocity of the fluid satisfying $u_\mu u^\mu =-1$, $\epsilon$ is the total energy density and $p$ is the gas pressure for a comoving observer. In the stress-energy tensor, the parts related to the dissipation due to the process like viscosity or the heat conduction is neglected. The corresponding redshift factor in the static set up is given by
\begin{align}
u_t=\sqrt{\frac{g_{tt}g_{\phi\phi}}{l^2 g_{tt}+g_{\phi\phi}}}\,,
\end{align}
where $l$ is the constant of motion in this context related to the specific angular momentum and energy, reads as
\begin{align}
    l & = \frac{L}{E}\,.
\end{align}
The equation $\nabla_\mu T^{\mu}{}_{\nu}=0$, where $\nabla_\mu$ denotes the covariant derivative with respect to the Levi-Civita connection, is then written as \cite{1978A&A....63..221A},
\begin{align}
-\nabla_\nu \ln u_t +\frac{\Omega \nabla_\nu l}{1-l\,\Omega} = \frac{1}{\epsilon +p}\nabla_\nu p\,.
\end{align}
where $\Omega$ is the angular velocity is defined by
\begin{align}
 \Omega=\frac{u^{\phi}}{u^{t}}\,.  
\end{align}
For a barotropic equation of state it turns out that this equation has solution when $\Omega=\Omega(l)$, then by integration
\begin{align}\label{maintori}
   - \ln{\frac{|u_t|}{|(u_t)_{\rm in}|}}+\int_{l_{\rm in}}^{l}{\frac{\Omega{\rm d}l}{1-\Omega l}}= \int_{p_{\rm in}}^{p}{\frac{{\rm d}p}{\epsilon +p}} =: W_{\rm in} - W,
\end{align}
where $(u_t)_{\rm in}$, $p_{\rm in}$ and $l_{\rm in}$ refer to the value of these quantities at the inner edge of the disk, and $W$ provides the equipotential surfaces topology of the disk.

Therefore, by choosing $l$ and explicitly $\Omega=\Omega(l)$, we can build the equipotential surfaces. There are different choices for specifying $l$; however, in the real astrophysical situations $l$ would be given by dissipative processes which are not known. Then, instead of prescribing this unknown dissipative processes, in the analytical approach one directly prescribes a model for the angular momentum $l$, either $l=\rm const.$ \cite{1978A&A....63..221A} or a non-constant angular momentum distribution e.g. \cite{2009A&A...498..471Q,2015MNRAS.447.3593W}.

In what follows, we consider a constant distribution of angular momentum to study the structure and the properties of these equilibrium configurations for different $\lambda$ parameters in the Born-Infeld teleparallel gravity. We compare the result to the Schwarzschild solution, which is its counterpart in GR.

\section{Results and discussion}\label{sec:RandD}

\begin{figure}
    \centering
    \includegraphics[width=0.4\textwidth]{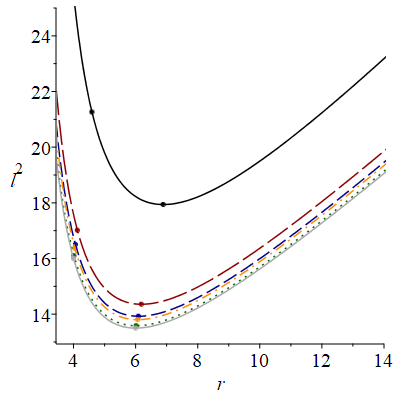}\quad 
       \caption{Specific angular momentum $l^2$ as a function of radius $r$ of circular orbits, for different choices of $\lambda$. From top to bottom line $\lambda=10,50,100,140,500,\infty$. The dots indicate the values of $l^2_{\rm mb}$ and $l^2_{\rm ms}$.}
    \label{fig:Lsq_circular}
\end{figure}

\begin{figure}
    \centering
    \includegraphics[width=0.8\hsize]{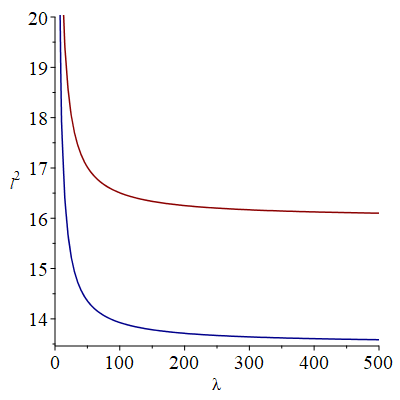}
    \caption{Specific angular momentum of the marginally bound orbit $l^2_{\rm ms}$ (lower blue line) and the marginally stable orbit $l^2_{\rm mb}$ (upper red line) as a function of $\lambda$.}
    \label{fig:Lmblms}
\end{figure}

\begin{figure}
    \centering
    \includegraphics[width=0.84\hsize]{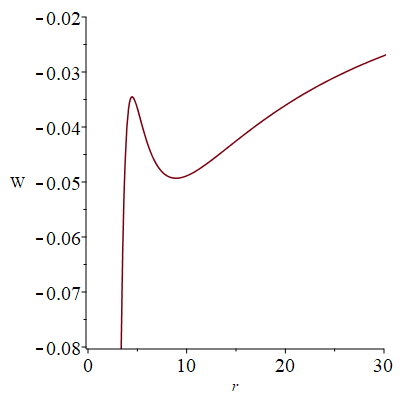}
    \caption{The function $W$ \eqref{maintori} in the equatorial plane for $\lambda=140$ and $l_0=(l_{\rm ms}+l_{\rm mb})/2$. Equipotential surfaces $W=\rm const$ corresponding to a bound disk can exist anywhere between the two extremal values. }
    \label{fig:EffPotW}
\end{figure}

For this analysis, we consider a constant specific angular momentum $l \equiv l_0$. In this case, the potential $W$ in \eqref{maintori} reduces to $W=\ln |u_t|$. It has extrema exactly at the positions where $l_0=l(r_0)$, where $r_0$ is the radius of a geodesic circular orbit in the equatorial plane. In Fig.~\ref{fig:Lsq_circular}, we therefore plotted $l^2$ as a function of the radius of circular orbit for different choices of $\lambda$, corresponding to the values presented in Table \ref{TTb}. Based on this investigation, several configurations can be constructed for different choices of $l_0$, namely, 
\begin{itemize}
    \item $l_0<l_{\rm ms}$: no disks possible 
    \item $l_{\rm ms} < l_0 < l_{\rm mb}$: bound disk structures are possible, one of them with a cusp
    \item $l_0>l_{\rm mb}$: the disk can not have a cusp
\end{itemize}
If $l=l_{\rm ms}$ we do not have a disk but a ring, and if $l=l_{\rm mb}$ the cusp is located at the marginally closed surface that just extents to infinity. The variation of these two angular momentum distributions $l_{\textrm mb}$ and $l_{\textrm ms}$ is shown in Fig.~\ref{fig:Lmblms} as a function of $\lambda$. In Fig.~\ref{fig:EffPotW} we show a typical example of the effective potential $W$ for the case $l_{\rm ms} < l_0 < l_{\rm mb}$. 

\begin{figure*}
\begin{tabular}{cc}
$\lambda=10$ & $\lambda=140$ \\
\includegraphics[width=0.35\textwidth]{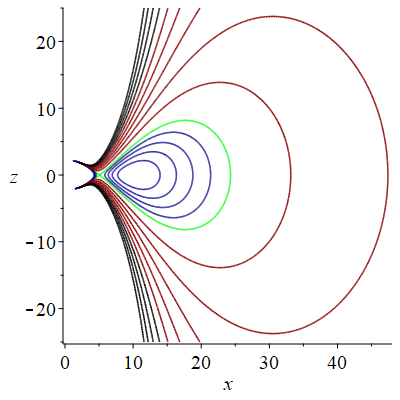}&
\includegraphics[width=0.35\textwidth]{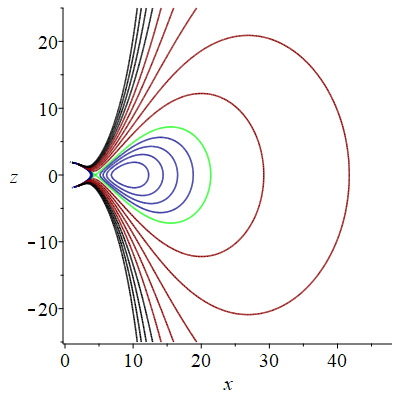}\\
$\lambda=500$ & $\lambda=\infty$\\
\includegraphics[width=0.35\textwidth]{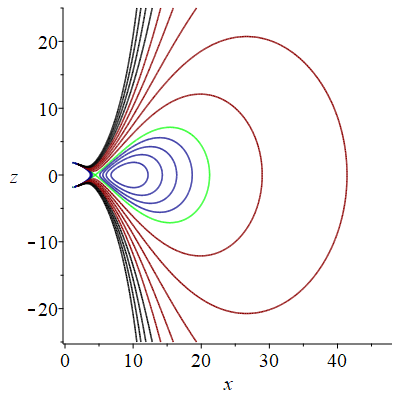}&
\includegraphics[width=0.35\textwidth]{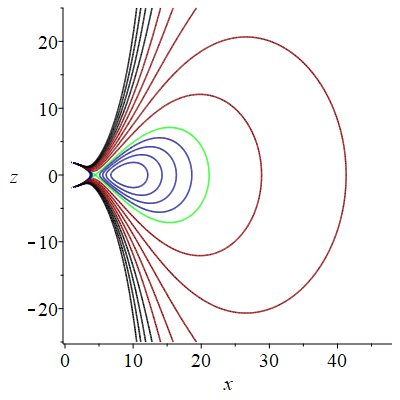}\
\end{tabular}
\caption{Equipotential surfaces for different choices of $\lambda$ and constant $l_0=(l_{\rm mb}+l_{\rm ms})/2$. The green line indicates the torus with a cusp corresponding to the maximum of $W$ on the equatorial plane. Blue lines indicate closed tori, red lines bound structures without inner edge, and black lines open surfaces.} \label{fig:equipotentials}
\end{figure*}

We now explore the dependence of the disk configurations on the spacetime parameter $\lambda$ for the particularly interesting cases where $l_{\rm ms} < l_0 < l_{\rm mb}$. To fix the specific angular momentum $l$ in this range in a way that enables a meaningful comparison between the different spacetimes, we choose to set it to the mean value $l_0=(l_{\rm ms}+l_{\rm mb})/2$ for each value of $\lambda$. For this choice, we show the disk configurations for different values of $\lambda$ in Fig.~\ref{fig:equipotentials}. In each plot, the green line indicates the equipotential surface corresponding to a cusp. Material from a disk filling this surface would then flow over the cusp and accrete into the central object, similar to a Roche lobe overflow (that is however not part of the analytical model). Blue lines inside the green line indicate possible bound disk structures without actual accretion. In general, a deeper analysis reveals that we have a larger disk size for smaller values of $\lambda$ as we anticipated from Fig. \ref{fig1}. However, by comparing the plots in Fig.~\ref{fig:equipotentials} we see that the influence of $\lambda$ on the shape of the disk is marginal for realistic values $\lambda \gtrsim 140$.

\begin{figure}
    \centering
    \includegraphics[width=0.4\textwidth]{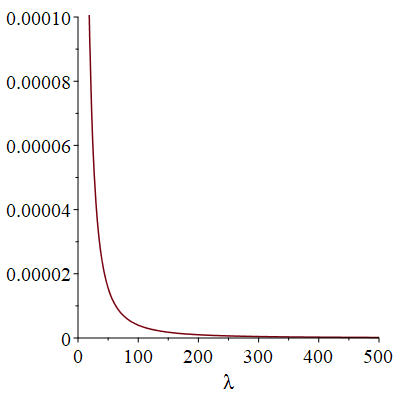}
    \caption{Variation of the maximal potential difference as function of $\lambda$. On the y-axis $\Delta W = \delta W_{\lambda} - \delta W_{\rm Sch}$ is plotted.}
    \label{fig:Potdiff}
\end{figure}

For further analysis of the influence of $\lambda$, let us consider the case of $l_0=l_{\rm mb}$, where the cusp is located on the marginally closed potential surface. Here the difference $\delta W$ between the potential value at the cusp and the center of the disk becomes maximal (Generally, assuming a specific equation of state for the system, this difference will translate into a pressure gradient). For the Schwarzschild case (corresponds to $\lambda=\infty$), this difference is exactly given by 

\begin{align}
\delta W = \frac{1}{2} \ln \frac{22}{9+5\sqrt{5}} \approx 0.0432.    
\end{align}
This difference generally increases for smaller $\lambda$. In Fig.~\ref{fig:Potdiff} we show the difference $\Delta W = \delta W_{\lambda} - \delta W_{\rm Sch}$. 

We again see that the influence of $\lambda$ on the potential differences, and therefore on the pressure gradients within the disk is very small, even for rather small values of $\lambda$. For comparison, in a maximal Kerr spacetime $\delta W = \frac{1}{2} \ln 3 \approx 0.55$ \cite{1978A&A....63..221A}. The influence of a rotation will therefore cover the influence of $\lambda$.

\section{Summary and conclusion}\label{sum}
Born-Infeld teleparallel gravity is one of the most successful f(T)-gravity theories so far, regarding the existence of non-perturbative solutions and comparison with experiments. In this article, we added a different piece to the puzzle of its viability and paved the way for further investigations by studying teleparallel accretion disks.

More precisely, we investigate the structure of equipotential surfaces of the analytical Thick accretion disk model in this setup. On our way to construct the Thick accretion disk of the Born-Infeld $f(T)$-gravity generalization of Schwarzschild geometry \eqref{eq:fTBImetric}, we studied the influence of the teleparallel modifications, parameterized by the theory parameter $\lambda$ on several further characteristics.

In particular, we found the expression of the Komar mass, which needs to be taken into account when the weak field limit of the solution is studied. It turned out that the PPN parameter $\gamma$ is identical to the one found in Schwarzschild geometry, but $\beta$ puts a bound on the teleparallel parameter of $\lambda\gtrsim 140$. Of course, studying and understanding the influence of the choice of tetrad on the PPN analysis of $f(T)$-gravity is an important future research direction, which became visible due to our analysis.

Furthermore, to construct the disk model we derived and analyze the effective potential, and explicitly the characteristic, marginally bound and marginally stable, orbits. Their dependence on the teleparallel parameter $\lambda$ was plotted in Fig.~\ref{fig1}. In fact, we saw a large difference only for much smaller values of $\lambda$. As byproduct we found the modification of the photon sphere. The first order modifications of these orbits are displayed in equations \eqref{rmb}, \eqref{rms} and \eqref{rph}, respectively.

Finally, we presented the panel of disk configurations for different values of $\lambda$ in Fig. \ref{fig:equipotentials}. In general, for larger values of $\lambda$ we have smaller disk configurations, which means the size of the disk is a monotonically decreasing function of $\lambda$. However, the influence of the teleparallel parameter $\lambda$ on the shape of the accretion disk is nearly invisible for $\lambda \gtrsim 140$. To quantify this effect we considered the difference in equipotential $\delta W$ between the cusp and the center of an accretion disk and compared it to the value in Schwarzschild spacetime, for $\lambda \gtrsim 140$. It turns out that  $\Delta W = \delta W_\lambda - \delta W_{\rm Sch} < 0.00002$, while for an extremal Kerr black hole we obtain $\Delta W = \delta W_{\rm Kerr} - \delta W_{\rm Sch} \approx 0.51$. Hence the influence of rotation on the accretion disk is way larger than the influence of the teleparallel parameter. For very slowly rotating black holes there exist degeneracies, in the meaning that it cannot be distinguished if $\Delta W$ originates from a teleparallel modification or a slow rotation. The achieved results are already exciting. Nevertheless, to discriminate between teleparallel gravity and GR effects considering rotation, the next step is to find a Born-Infeld teleparallel gravity solution which generalizes Kerr geometry whose nonrotating limit coincides with the spherically symmetric solution studied in this work.


\begin{acknowledgments}
The authors thank the anonymous referee for the remarks which promoted the paper. C.P. is funded by the Deutsche Forschungsgemeinschaft (DFG, German Research Foundation) - Project Number 420243324. C.P., E.H., and S.F. acknowledge the research training group GRK 1620 "Models of Gravity" funded by the Deutsche Forschungsgemeinschaft (DFG, German Research Foundation) and the excellence cluster QuantumFrontiers funded by the Deutsche Forschungsgemeinschaft (DFG, German Research Foundation) under Germany’s Excellence Strategy – EXC-2123 QuantumFrontiers – 390837967. S.B. is supported by JSPS Postdoctoral Fellowships for Research in Japan and KAKENHI Grant-in-Aid for Scientific Research No. JP21F21789. S.B. also acknowledges the Estonian Research Council grants PRG356 ``Gauge Gravity"  and the European Regional Development Fund through the Center of Excellence TK133 "The Dark Side of the Universe". 
\end{acknowledgments}

\bibliographystyle{unsrt}
\bibliography{TPAcc}

\end{document}